# Comparison of Wireless Standards-Setting
## --United States Versus Europe.

Draft Paper (09/01)
Zixiang (Alex) Tan
Syracuse University
ztan@syr.edu

## 1. Introduction

Standards-setting can generally be divided into three types. *Proprietary standards* are developed by an individual firm or firms through the market process. Standards institutions set up *institutional standards* through their consensus decisions. Governments intervene and establish *governmental standards* when they are deemed necessary.

When decisions about developing standards are left to individual firms, there are many advantages including a flexible response to market evolution, accommodation to rapid technology change, and avoidance of costly coordination. However, the market process does not necessarily lead to compatible standards, especially when there is no clear market dominator. The market may pick up a wrong or technically inferior standard. In addition, if the market-produced standard is a proprietary design of a particular firm, the firm may assert some monopolistic practices through the holding of patents and other Intellectual Property Rights (IPR), which puts other firms at disadvantage and results in efficiency losses to society.

Standards Institutions generally facilitate better communication among market participants, which may discourage early and/or primitive incompatible standards to emerge in the market. Better involvement of all market participants tends to develop higher-quality standards. Moreover, an institution-developed standard is a public good which prevents the emergence of monopoly on standards. However, expenditures on institutional standards-setting can lead to diminishing returns, or even be counter-productive because institutional standards usually take longer to produce and are slower to respond to technology development. Competing institutions or competing groups within one institution may promote competing standards.

Government involvement is often called for when market competition and the institutional process seem in danger of producing multiple incompatible standards. Government agencies can specify compulsory standards to avoid incompatibility. They can also work closely with institutions to produce a de facto single standard. In addition, governments can tie their policy on standardization closely with their industrial, trade, and/or regulatory policies. However, many exogenous factors handicap a government's effective involvement. As a bureaucratic organization, government often fails to respond to the dynamics of technology development and consumer demand, and to pick up the "right technology". In summary, all three standards setting methods are imperfect.

Standardization of world-wide wireless communications systems is caught in this dilemma. Rapid technology development and the changing demands of wireless systems make the dynamics of proprietary standards-setting desirable. However, the cost of too many incompatible standards, lack of dominant market leaders, and danger of proprietary technology monopolies all point to the need to seek help from institutional standards-setting. Concerns about global technology competition and trade, furthermore, justify intervention from governments. Because of these factors, different models of standards-setting for wireless systems have emerged in different countries.

Globally speaking, standards-setting in wireless systems is centered in the United States and Europe. In the U.S., the FCC has decided not to specify any mandatory standard since 1980s. The Telecommunications Industry Association (TIA), as a typical industry association, has taken over the standard-setting responsibility. TIA's role is to create a forum for participants to exchange information and work out detailed specifications. Once a particular proposal gains solid industry support, it is published as a voluntary standard. This process leads to an institutional standard-setting environment favoring competing standards sponsored by temporary industry alliances. The U.S. standards-setting model could be best characterized as an institutional one supported by market competition.

In Europe, the liberalization of the telecommunications industry and the European Union's (EU) political ambition of creating a single European market have led to the adoption of a new standards-setting model. A Pan-European institution, European Telecommunications Standard Institute (ETSI), has emerged as the single entity to create standards for the European market. Although ETSI's standards are in the form of recommendations, the EU has developed some political and economic mechanisms to coordinate and implement recommendations from ETSI among its member nations. Standards from ETSI are often made "mandatory" or "quasi-mandatory" by the EU. The European model could be characterized as government-imposed institutional standards-setting.

This study first discusses why two different models have emerged in the two regions. It then examines outcomes and implications of the two models, including impacts on domestic service deployment, global trade competition, technology innovations, and strategies of multinational corporations. The goal is to conduct an empirical comparison between the two contrasting models. Given the high likelihood that both regions will continue their current standards-setting model, the findings will be significant to help us understand and predict the future scenario of wireless standards-setting in the world.

## 2. Standardization in Global Wireless Communications

### 2.1. Three Generations of Wireless Communication Technologies in the World

Wireless communications is referred as terrestrial based public wireless systems for this paper. Up until now, wireless communications can be categorized into three generations. The first-generation (1G) networks were developed and installed in the early 1980s. All the 1G systems used analog technology that relied on Frequency Division Multiple Access (FDMA) methods to create multiple radio channels for multiple users. The 1G standards include the American Mobile Phone System (AMPS) in United States, Nordic Mobile Telephone (NMT) in Europe, Total Access Cellular System (TACS) in Britain, NTT system in Japan, and others.

Second-generation (2G) standards were developed and installed in the early 1990s. These systems have shifted to digital technology, primarily using Time Division Multiple Access (TDMA) methods to create multiple access channels for subscribers. Some 2G systems have deployed Code Division Multiple Access (CDMA) technology, which has further improved system capacity and spectrum efficiency. Generally speaking, 2G standards have achieved significant improvements in system capacity, service quality, and information security among other features, compared with 1G systems. However, 2G systems continue being voice-communication focused. Popular 2G standards include the Global System for Mobile (GSM) from Europe, US TDMA and US CDMA from the United States, and Pacific Digital System (PDS) from Japan. 2G systems have successfully brought the wireless voice services to mass users.

The third-generation (3G) standards are currently being developed under the umbrella of ITU's standard - the International Mobile Telecommunications 2000 (IMT-2000). The dominant standards have adopted Code Division Multiple Access (CDMA) technology to create access channels for users while the legacy TDMA systems such as GSM and TDMA/136 will be migrated to provide 3G applications. The major improvements targeted by the 3G standards include higher spectrum efficiency, better compatibility among different standards and a larger bandwidth for high-speed data communications. The ultimate goal is to bring the Internet and multimedia services to the mobile users (Nilsson, 1999).

## 2.2. Three Generations of Standardization Models in the World

The 1G standards were mainly products from a sponsored environment (David & Greenstein, 1990) and featured with a national endorsement. It was often an old national regime for a domestic service monopoly, an equipment supply champion(s), or a combination of the two to make their own technical standards a de facto result from a sponsored market or administrative process. In the case of AMPS standard in the U.S., AT&T created inducements for other decision-makers to adopt its particular set of technical specifications and to become part of a national AMPS network. The Japanese standard was mainly sponsored by its service monopoly the NTT. The NMT standard was developed and sponsored by the dominant Nordic equipment suppliers, Ericsson and Nokia (Kano, 2000). On the other hand, many other nations did not have the technical capacity to develop their own standards. The choice for them was to pick up a particular standard developed in a domestic market.

The 2G standardization shifted away from the 1G models. Important factors contributing to the changes included the global trend toward telecommunications liberalization and deregulation, the changing national policy on standards, and the changing economic and political environment as well as advances in technology. Overall, the making of 2G standards is dominated by three models: a ***liberalized U.S. environment*** favoring competing standards sponsored by temporary alliances; a ***highly politicized European model*** of promoting a Pan-European single standard in wireless communications; and a ***national sponsored Japanese model*** to formulate national domestic standards.

The 3G standardization has evolved to a new model with the involvement of the International Telecommunications Union (ITU). ITU's standard, IMT-2000 (International Mobile Telecommunications 2000), works as an umbrella to facilitate compatibility and implementation (Nilsson, 1999). Under the umbrella is the cooperation and competition among standards sponsored by institutions, corporations, all kinds of alliances, and even governments around the world. The competition is mainly among a single European standard and competing U.S. standards with Japan giving up its efforts to sponsor its own 3G standard. While China and South Korea have joined the competition by sponsoring their own standards, other countries again have to pick up their own favorite ones from the competition.

## 3. Different Standardization Models: U.S. Vs. Europe

*3.1. The U.S. Model*

The "cellular" concept and the frequency re-use in small cells were originated from AT&T's Bell Laboratory in 1947 (McDonald, 1979), which has provided a fundamental basis for all the wireless systems in the world. Based on the cellular design, AT&T then developed a new public wireless phone system called AMPS (American Mobile Phone Systems). AMPS was technically ready to roll-out to the market in the early 1970s. The commercialization process was delayed by frequency allocation, regulatory policy, and industry politics. The first developmental system trial was granted in 1977 and actually was conducted by Illinois Bell in Chicago in 1978. Subsequently, the Chicago cellular system began the first AMPS commercial operation by AT&T on October 13, 1983.

Despite the delay of the service introduction, AMPS was the only standard for the first-generation US public mobile system. Two major factors made it possible. First, only AT&T had conducted serious standards setting activities in the United States until the early 1980s. The AMPS was thus the only standard commercially available. No competing standards existed. This was a perfect result of AT&T's monopoly in standards setting. Secondly, the FCC's licensing process specified AMPS as the unified, nationwide standard. This model was consistent with those in other countries: a single domestic standard under sponsored environment and with a national endorsement.

In the United States, AT&T's monopoly both in telecom services and in standards setting was dismantled by the 1982 MFJ divestiture (Besen & Saloner, 1988; Wallenstein, 1990). The FCC decided not to specify any mandatory standard for the 2G public mobile phone systems while recognizing the technology dynamics and the FCC's limited resources and expertise. License holders granted by the FCC could utilize whichever standard according to their business considerations and strategies. The TIA (Telecommunications Industry Association), as a typical industry association, took over the standard-setting responsibility. The TIA does not recommend which standard is technically superior or inferior. Its role is to create a forum for participants to exchange information and work out the detailed specifications. Once the work gains solid industry support, it is published as a voluntary standard. This overall change leads to a *liberalized environment* favoring competing standards sponsored by temporary alliances. The TIA has published two competitive 2G standards promoted by different corporation alliances.

Starting from 1985, CTIA (Cellular Telephone Industry Association) launched a systematic evaluation of various technological alternatives. This was endorsed by cellular operators and major equipment manufacturers including Motorola, AT&T, Nortel, Ericsson, and IMM. TDMA, CDMA, N-AMPS, and E-TDMA were among the candidates. In 1989, CTIA members voted TDMA as the standard for the 2G mobile systems mainly based on its commercial readiness and availability. CTIA then requested the EIA/TIA to set up the detailed technical specifications for the TDMA standard.

In April 1992, EIA/TIA published the TDMA standard as IS-54 and validation tests were undertaken in 1991. However, the deployment of TDMA proceeded slowly. CDMA, which was considered too far from commercial deployment in the early 1990s, has rapidly and successfully caught up. Since its initial discussion with PacTel in February 1989, Qualcomm successfully persuaded CTIA to bring the CDMA technology to EIA/TIA. In July 1993, EIA/TIA published its CDMA standard as IS-95. Standardization in the US wireless industry shifted from its 1G of a single standard approach to its 2G of multiple, competitive standards.

### 3.2. The European Model

The first European analog system, NMT450 (Nordic Mobile Phone 450), was developed in Sweden. The four continental Nordic countries (Denmark, Finland, Norway and Sweden) first introduced a common 450Mhz cellular system in 1981-2. Other European countries followed suit rapidly. By 1990, there were already 3 million subscribers in total. However, incompatible technical standards were utilized. There were in total six major incompatible standards in Europe - NMT 450, NMT 900, TACS, C450, Radiocom 2000 and RTMS.

NMT450 and NMT900 were developed and first used in the Nordic countries. They were the most widely adopted standards in Europe by 1990. In addition to Finland, Norway and Sweden, Denmark, Netherlands, Belgium and Luxenburg also adopted NMT system. TACS (Total Access Communication System) was a British standard, which was a modified version of the US AMPS system. By 1990 TACS had gained a substantial market share outside Britain, including Ireland, Greece, and Spain. TACS systems had also been adopted in many countries outside Europe, including Hong Kong, Singapore, and China. Germany developed its own C450 standard, which was exported to Austria and Portugal. Radiocom 2000 was designed in France and deployed only in France. RTMS was an Italian standard, which was only installed in Italian market.

The European market was fragmented by incompatible standards. Service stopped at a nation's borders because of this incompatibility. Roaming function could only be engaged in a few limited geographical areas where the same standards were adopted. European manufacturers had not realized a Europe-wide economy of scale, which hindered a rapid decrease in equipment prices. Efforts in marketing, maintenance services, and research and development could not be coordinated in a European level.

In Europe, telecommunications liberalization and EU's (European Union) political ambition of creating a single European market led to the adoption of a new standardization model in telecommunications in general and wireless communications in particular. A Pan-European institution, European Telecommunications Standard Institute (ETSI), emerged as the single entity to create standards for European market. Although ETSI's outcome is in a form of recommendations, the EU has built up some political and economic mechanism to coordinately implement recommendations from ETSI among its member nations. Standards from ETSI often become "mandatory" or "quasi-mandatory" by the EU (Hawkins, 1992). For the 2G mobile systems, EU's strategic involvement included its implementation timeline for member states, common frequency bands, and single standard requirement (Tan, 2001). This leads to a ***highly politicized European model*** of promoting a Pan-European single standard in mobile communications.

A scenario of a Pan-European wireless telephone system was first proposed by five PTT Administrations from the Netherlands, Denmark, Finland, Norway and Sweden in 1982 (Cattaneo, 1994). The proposal was accepted. The study group GSM (Group Special Mobile) was formed to study the specifications of a possible Pan-European standard in the same year. Because of the conflicts and vested interests among European nations and major corporations, a balanced standard that could be acceptable to all concerned interest groups was outlined in February 1987. The outline agreed to make the compromise: adopting a digital instead of an analog technology as the pan-European standard. However, it was a new digital standard but not Alcatel-SEL's proprietary one which was favorable to the German and French PTTs.

On the political side, in May 1987 ministries from the four largest markets - France, Germany, Italy, and the UK - agreed to accept the basic GSM standard proposed by the GSM working group. The four countries actually drafted a Memorandum of Understanding (MoU) to commit themselves to the elaboration and deployment of the GSM standard. The MOU laid down a series of milestones for the remaining technical and regulatory issues. It also planned to start commercial GSM services in June 1991. By September 1987, 13 other countries joined the effort, increasing the signatory countries of the MoU to 17. Meanwhile, the EC used its political influence on its member countries to guarantee GSM's success. On June 25, 1987, EC's Council of Ministers adopted a Recommendation and a Directive that played a crucial role for Europe to achieve consensus agreement about public mobile communication.

GSM began its implementation stage in 1988 after the above-mentioned key events in 1987. The detailed technical specifications were developed by the GSM working group. The GSM system finally hit the market in 1992. By 1995, there were already 50 networks in operation in the world. This achievement put an end to the incompatible standards in Europe. The European public mobile market has evolved to a single and unified standard.

3.3. U.S. Vs. Europe: Different Paths

The U.S. and Europe have apparently adopted very different policies toward standardization in wireless communications. They started from a same base when developing 1G standards: a single domestic standard sponsored by a national service or manufacturing champion and endorsed by the national regulator. This model has been totally changed both in the U.S. and in

Europe because of the changing market environment, the deregulation, and the technology dynamics.

The U.S. has moved toward to market competition for standardization. The core of this policy is to promote the coexistence of multiple competing standards. It is left to the market to determine the final outcome of the competition among different standards. Europe has moved to a government intervened institutional coordination. The ultimate European goal is to agree on a single standard and to deploy the same standard across the Europe. This comparison between the U.S. and the Europe is illustrated in Figure One. These approaches for the U.S. and the Europe seem to continue in the foreseeable future.

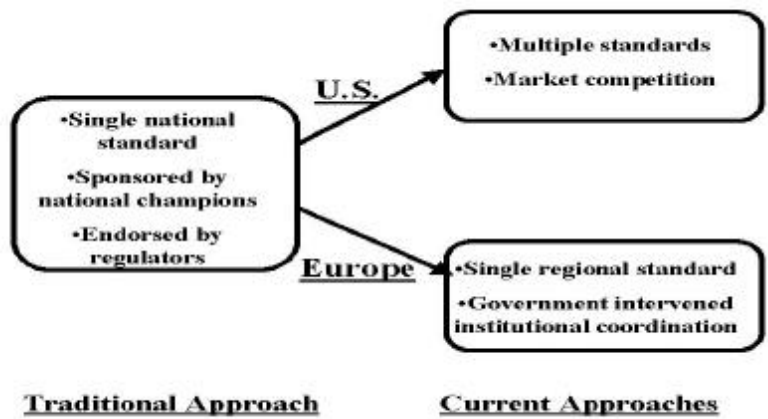

Figure One: The Divergence of U.S. and European Paths.

## 4. Impacts of the U.S. and European Approaches

### 4.1. Standards in the domestic markets

Standards competition in any domestic market is shaped by several factors, including standardization policy, licensing policy, and service providers' strategies. In the U.S. market, market oriented standardization policy has led to the emergence of competing TDMA standard and CDMA standard. In addition, current U.S. licensing policy allows service providers to choose whatever standard they deem appropriate, which has opened the door for Europe's GSM standard to land in U.S. market. As the world moves toward 3G standards, the US CDMA standard would evolve to CDMA200 and the GSM standard would evolve to W-CDMA. The current coexistence of three major standards in the U.S. market will most likely lead to two competing standards in its 3G market.

The European standardization policy has strongly favored a single European standard. This is implemented through institutional coordination in telecommunications industry and the political intervention of European Union and national governments. Service providers do not have many incentives to switch to a non-European standard since they might be isolated in the market. Europe's 3G standard, W-CDMA, has clearly geared to evolve from current GSM standard. This technical effort, together with European Union's continuous policy of having one single

European 3G standard, has pointed to a future single standard in European market. Figure Two illustrates the available standards in U.S. and Europe markets.

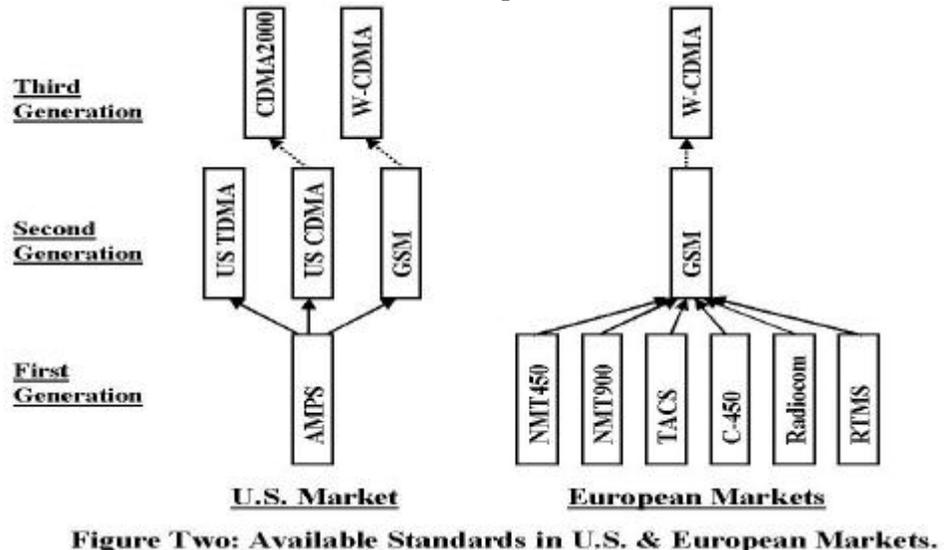

Figure Two: Available Standards in U.S. & European Markets.

**4.2. Service competition in the domestic market**

The outcomes of different standardization policies have influenced the competition among service providers. In U.S. market, one of the most relevant strategies is *for one operator to differentiate from other competitors by choosing a particular standard for its system*. Operators tend to choose US TDMA standard when they have had a large 1G (AMPS) network. They often claim to provide high service reliability, complete geographic coverage, and smooth immigration from 1G services. AT&T Wireless and others are among these operators. Operators, that choosing UC CDMA standard, often play the high capacity of the CDMA technology over the TDMA technology. The current CDMA standard also gives these operators the credit to upgrade to future generation CDMA standards. Sprint PCS has used its chosen CDMA standard to claim its superiority over TDMA operators. The newly emerged GSM operators in U.S. market have linked their advantages to the GSM's popularity in the global market. This global popularity allows GSM users great compatibility across many nations in terms of global travelling and new applications such as Short Message Services (SMS). In most markets in the United States, the competition is among three major carriers with three different standards: US TDMA, US CDMA, and GSM. In addition, the unsettled competition among three standards has added uncertainty to who would win in the market, which has both delayed the network deployment and expansion and caused some service providers to switch from one standard to another. AT&T Wireless has been reported to built up its GSM network on top of its existing US TDMA network to harvest the general benefits of GSM technology in the global market.

In European market, a single same standard pushes competition among service providers to different pricing schemes, after-sale supports, and network facility expansion and improvement. Number portability can be effectively implemented since customers do not need to switch their terminal equipment when switching from one service provider to another. Services such as Short Message Services (SMS) can be implemented across country or even region. Technology differentiation is not an alternative competitive strategy in European countries. Service operators, equipment manufacturers, and users were all certain that every country would use the same standard, GSM, in Europe. This certainty, to a large extent, resulted in rapid installation and expansion of 2G mobile systems in Europe.

**4.3. Standards competition in the global market**

The European Union's policy toward a Pan-European single standard led to the birth of GSM standard, which gained a large subscriber base in Europe. Being the first available digital standard in the world and having a significant installed base in Europe, GSM had been favorably received by standard-adopting countries in the world. Service operators across the world recognize that GSM has already started as a bandwagon in Europe. It is relatively low risky to go with the GSM standard, compared with the unsure US TDMA and US CDMA standards. As a result GSM standard has rolled out rapidly in the global market. As shown in Figure Three, there were over 120 countries that had chosen GSM standards in January 1999. However, there were 12 countries choosing US TDMA standards and 24 choosing US CDMA. U.S. standards have lost ground in the global market to the European standard.

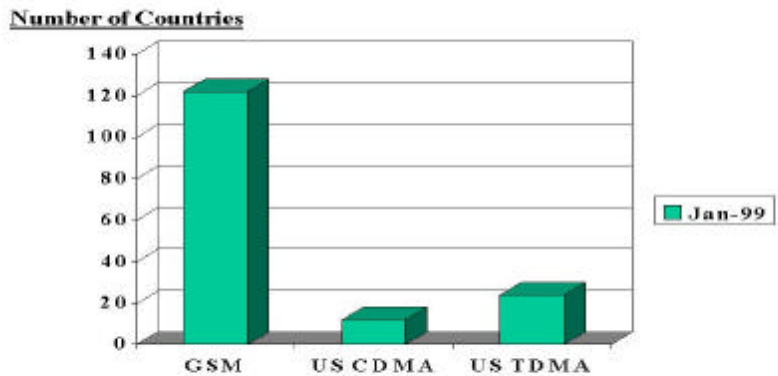

Figure Three: 2G Standards Adoption Distribution in January 1999.
Source: GSM Association.

On the opposite in the U.S. market, standards competition among TDMA and CDMA standards makes none of them a clear winner. Even some U.S. domestic carriers have had hard time to make up their mind regarding which standard they should adopt. This uncertainty failed to convince other countries to adopt U.S. digital standards, even though CDMA standard appeared to promise some technical advantages over the European GSM standard. The result is a rapid global deployment of GSM standard, together with a slow deployment of US TDMA and US CDMA standards in the global market, as shown in Figure Four.

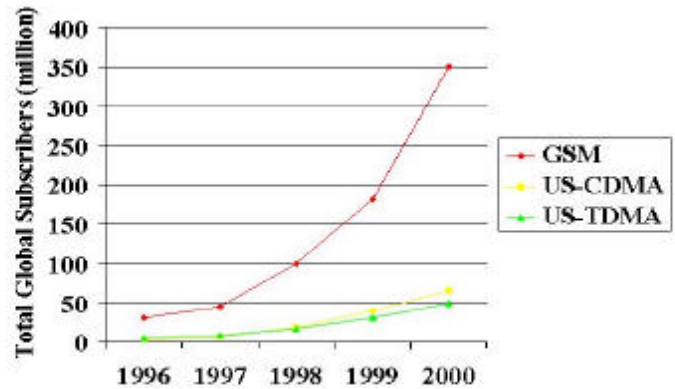

Figure Four: Total Global Subscribers Among Different Standards
Sources: GSM Association.

## 4. Conclusions

While the U.S. and Europe have chosen very different standardization policies regarding public wireless communications, the impacts are significant. The market competition approach in the U.S. leads to multiple competitive standards, which coexist in the U.S. market. Service providers have largely tried to differentiate from each other by choosing different standards, which results in associated benefits as well as drawbacks. This coexistence failed to convince other countries to commit to U.S. standards.

On the opposite, the government-intervened standardization policy in Europe leads to a single unified standard, GSM. While deploying a same GSM standard, European service providers focused their competitive energy on different pricing schemes, after-sale supports, and network facility expansion and improvement. A solid installed base of the European GSM standard pushes the rest world to jump into its bandwagon, which leads to a rapid growth of GSM subscribers in the global market.